\documentclass[12pt]{article}
\usepackage{latexsym}

\begin{document}

\newcommand{\be}{\begin{equation}}
\newcommand{\ee}{\end{equation}}
\newcommand{\<}{\langle}
\renewcommand{\>}{\rangle}
\newcommand{\reff}[1]{(\ref{#1})}

\title{Series expansion of the quark determinant in the number of quark-antiquark pairs} 
\author{
  { Fabrizio Palumbo~\thanks{Work supported in part by the 
European Community's Human Potential Programme under contract
HPRN-CT-2000-00131 Quantum Spacetime}}            
\\[-0.2cm]
  {\small\it INFN -- Laboratori Nazionali di Frascati}  \\[-0.2cm]
  {\small\it P.~O.~Box 13, I-00044 Frascati, ITALIA}          \\[-0.2cm]
  {\small Internet: {\tt palumbof@lnf.infn.it}}     
   }
\maketitle
\date{April 15, 1997}
\thispagestyle{empty}   


\begin{abstract}

We propose a formulation of the QCD partition function which leads to a series 
expansion for the  quark determinant in any given baryonic sector. The r-th term gives the  
gauge-invariant contribution of the valence quarks plus r quark-antiquark pairs.
This expansion can be used to investigate any  baryonic sector, 
starting from the nucleon up to high baryonic densities. 

\end{abstract}

\clearpage

\section{Introduction}

A clearcut separation of the valence and sea quark contributions to the QCD partition
function might help in the study of a number of issues, starting with the spin content
of the nucleon~\cite{Math} up to QCD at finite baryonic density~\cite{Sand}. The 
identification of the different contributions can be achieved by means of an expansion  of
the partition function with  respect to the number of quark-antiquark pairs, which  is the
goal of the present paper. But the practical usefulness of such an expansion depends on the
number of terms necessary to get the desired accuracy. 

In the case of the nucleon, for instance, only a few quarks of the sea are  expected
to contribute significantly. If the partition function is  split in the corresponding 
terms only a few will be important.

At first sight the case of QCD at nonzero baryon density appears quite different, 
because in principle one is interested in the termodynamic limit. To investigate this limit
an infinite number of terms would obviously be necessary, making our expansion of little use.
In actual numerical simulations, however, the physical volumes accessible are of the order
of a  few $fm^3$. If we consider that  normal nuclear density is about $3\,quarks /
6\,fm^3$,  with 6 valence quarks we are  already at about twice the ordinary nuclear density,
and it is reasonable to expect that the number of important quark-antiquark
pairs be comparable to the number of valence quarks at least at moderate density and
temperature. 

To our knowledge there exists no clearcut separation of the contributions
we are discussing. To tackle this problem we need, to start with, a
formulation of the partition function in a given baryonic sector. One such formulation
has been constructed by introducing an imaginary chemical 
potential in the grand canonical partition function and by projecting in a given 
baryonic sector by a Fourier transformation~\cite{Mill},~\cite{Kacz}. But the result is not
very transparent with respect to the distinction of the different contributions we are
interested in.

We consider here a different form of projection, which leads to a series expansion 
for the quark determinant. The terms of this expansion have the desired physical meaning: 
The r-th term gives the gauge-invariant contribution of the valence quarks plus r
quark-antiquark pairs. In view of the above considerations, we hope that our expansion 
will have a reasonable convergence in present numerical simulations.

The technique we employ is fairly general, requiring only the knowledge of the transfer 
matrix in the Fock space of the quarks. But in this paper we restrict ourselves to
the case of Wilson fermions. We hope to be able to treat the Kogut-Susskind fermions
in a future work.

We outline the derivation of the series expansion in Sec.2 focussing on 
the strategy, leaving out many details of the calculations for 
Sec.3. In Sec.4 we give our conclusions.

\section{The series expansion of the quark determinant in a given baryonic sector}

In this Section we outline the steps leading to a series expansion of the 
quark determinant in a given baryonic sector. As already said the presentation will 
be somewhat schematic.

The quark determinant in the absence of any condition on the baryon number is
\be
det \, Q = \int [d \overline{\psi} \psi] \exp S_F.
\ee
$S_F$ is the quark action and $Q$ the standard quark matrix which will be spelled out
later. Our
strategy is to write $det \, Q $ as the trace of the transfer matrix acting
in the quark Fock space, impose the restriction to a given baryonic sector,  and then rewrite
the trace as the determinant of a modified quark matrix. The round trip is done by  mapping
the Grassmann algebra generated by the quark fields into the Fock space  following the
construction of Lu\"scher ~\cite{Lusc}. But while his paper is based on the mere existence
of the map, for us it is essential a concrete realization by means of coherent states.

 In the transfer matrix formalism the (euclidean)
time is treated differently from the spatial coordinates. So we
must also treat it differently, and we assume a convention of summation
over intrinsic quantum numbers and spatial sites only. These have coordinates
${\bf n}$, while the time will be denoted by $n_0$.
The link variables $U_{\mu}$ are matrices in color and have spatial matrix elements
\begin{eqnarray}
\left( U_0(n_0) \right)_{{\bf m},{\bf n}} &=& U_0({\bf m},m_0) \delta_{{\bf m}, {\bf n}}
\nonumber\\
\left( U_j(m_0) \right)_{{\bf m},{\bf n}} &=& U_j({\bf m},m_0) 
\delta_{{\bf m}+ {\bf j}, {\bf n}}.
\end{eqnarray}
The quark field $\psi$ carries Dirac, color and flavor indices, denoted altogether by
$\alpha$, and space-time label $n$.
We will use with the same convention Grassmann variables  $x$'s and $y$'s  and quark
creation and annihilation operators $\hat{x}^+,\hat{x},\hat{y}^+,\hat{y}$. But
unlike Grassmann variables, the creation and annihilation operators do not carry a temporal
index. Thus according  to our convention
 \begin{eqnarray}
{\overline \psi}(n_0) \psi(n_0) &=& 
\sum_{{\bf n},\alpha}{\overline \psi}_{{\bf n},\alpha}(n_0) \psi_{{\bf n},\alpha}(n_0)
\nonumber\\
\hat{x}^+ \hat{x} &=& \sum_{{\bf n},\alpha}\hat{x}_{{\bf n},\alpha}^+  \hat{x}_{{\bf
n},\alpha},
\nonumber\\
\hat{x}^+ x(n_0) &=& \sum_{{\bf n},\alpha}\hat{x}_{{\bf n},\alpha}^+ 
x_{{\bf n},\alpha}(n_0)
\end{eqnarray}
and the quark action is
\begin{eqnarray}
S_F &=&  \sum_{m_0,n_0}\overline{\psi}(m_0) Q(m_0,n_0) \psi(n_0)
\nonumber\\
&=& \sum_{m,\alpha,n,\beta}\overline{\psi}_{{\bf m},\alpha}(m_0) 
Q_{{\bf m}\alpha,{\bf n},\beta}(m_0,n_0) \psi_{{\bf n},\beta}(n_0).
\end{eqnarray}

Let us come back to the evaluation of the quark determinant in a given baryonic
sector. 

The first step is to write the unconstrained determinant as a trace in Fock space
\be
det \, Q= Tr \, \hat{\cal{T}}.  \label{trace}
\ee
$\hat{\cal{T}}$ is the transfer matrix whose definition will be given in the next Section.
The second step is to impose the restriction to a sector with baryon
number $n_B$ by inserting in the trace the appropriate 
projection operator $P_{n_B} $
\be
det \, Q|_{n_B} = Tr \left( \hat{\cal{T}} P_{n_B} \right).
\ee
To rewrite the above expression as a determinant we choose coherent states~\cite{Nege} 
as a basis 
\be
|x> = |\exp( - x \,\hat{x}^+ ) >,\,\,\,|y> = |\exp( - y \, \hat{y}^+ )>.
\ee
 In such a basis the trace takes again the form of a Berezin integral 
\begin{eqnarray}
det \, Q|_{n_B} &= & \int [dx^+ dx\, dy^+ dy ]
\exp \left(- x^+x-y^+y \right)
<x,y|\hat{\cal{T}} P_{n_B}|-x,-y>  \label{Berezin}
\nonumber\\
\end{eqnarray}
which will be expressed in terms of the determinant of a modified quark matrix. 

So far for the strategy. Now we show how we get a series expansion, what is its
the physical interpretation and what is the final result. 

The kernel $<x,y|\hat{\cal{T}} P_{n_B}|-x,-y> $ has the integral form
\begin{eqnarray}
<x,y|\hat{\cal{T}} P_{n_B}|-x,-y> &= & \int [dz^+ dz\, dw^+ dw]
\exp \left(- z^+z-w^+w \right) \cdot
\nonumber\\
 & & <x,y|\hat{\cal{T}}|z,w><z,w| P_{n_B}|-x,-y> . \label{convol}
\end{eqnarray}

The kernel $<x,y|\hat{\cal{T}}|z,w>$ will be given in the next Section, while 
that of $P_{n_B}$ is immediately calculated 
\begin{eqnarray}
< z,w|P_{n_B}|-x,-y> &=& \sum_{r=0}^{\infty}(-1)^{n_B}{1 \over ((n_B +r)! r!)^2} \cdot
\nonumber\\
 & &  <(\hat{y} w^+)^r (\hat{x} z^+ )^{(n_B+r)} 
(x \hat{x}^+ )^{n_B+r} (y \hat{y}^+)^r >.
\end{eqnarray}
 Since ${\hat x}^+ , {\hat y}^+$ are creation operators of quarks and antiquarks
respectively, we see that
the r-th term of this  series gives the gauge-invariant contribution of $n_B$ valence 
quarks plus $r$ quark-antiquark pairs. 
 
Needless to say, for $n_B=0$, $det \, Q|_{n_B}$ does not reduce to the
unconstrained determinant: Indeed also baryonic states are present in the
unconstrained
determinant, while they are not in $det \, Q|_{n_B=0}$. In QCD at nonvanishing
temperature it makes a
difference  whether we impose or not the condition $n_B=0$. 
In view of the
relatively low value of the critical temperature with respect to the nucleon mass, however,
we do not expect significant effects from this restriction unless we go to exceedingly
 high tempreatures.

Let us now proceed to derive our final result. By evaluating
the vacuum expectation values appearing in the last equation we express the kernel 
of the projection  operator in terms of Grassmann variables only
\begin{eqnarray}
< z,w|P_{n_B}|-x,-y> = \sum_{r=0}^{\infty}(-1)^{n_B}
{1 \over (n_B +r)! r!} (z^+ x)^{n_B+r} (w^+ y)^r.
\end{eqnarray}
To evaluate the integral of Eq.\reff{convol} we rewrite the above
equation in exponential form
\begin{eqnarray}
< z,w|P_{n_B}|x,y> = \sum_{r=0}^{\infty}{1 \over (n_B +r)! r!}
{\partial^{n_B+r}
\over 
\partial j_1^{n_B+r} } {\partial^r \over \partial j_2^r } 
\exp(-j_1 z^+x - j_2 w^+ y ) |_{j_1=j_2=0}.
\end{eqnarray}

The integrals of Eqs.\reff{Berezin},\reff{convol} are gaussian and we get the
constrained determinant in terms of the determinant of a modified quark matrix
\be
det \, Q|_{n_B} =  \sum_{r=0}^{\infty}{1 \over (n_B +r)! r!} {\partial^{n_B+r} \over 
\partial j_1^{n_B+r} } {\partial^r \over \partial j_2^r } 
\left( Q + \delta Q \right) |_{j_1=j_2=0}.
\ee
The variation of the quark matrix is
\begin{eqnarray}
\delta Q & =& K \left[ (j_1-1) ( 1+\gamma_0) \, d(N_0-1) \, U_0(N_0-1) \, T_0^{(+)} + \right.
\nonumber\\
 & &   \left.(j_2-1) ( 1 - \gamma_0) \, d(N_0) \, T_0^{(-)} \, U_0^+(N_0-1) \right.
\nonumber\\ 
   & &  \left. + (j_1 j_2 -1) \, d(N_0) \, C(N_0) \, ( 1 + \gamma_0) \right],  \label{delta}
\end{eqnarray}
where
\begin{eqnarray}
\left( T^{(\pm)}\right)_{m,n} &=& \delta_{m_0 \pm 1, n_0} \delta_{{\bf m}, {\bf n}},
\nonumber\\
\left( d(N_0) \right)_{m,n} &=& \delta_{m_0 , N_0} \delta_{m_0 , n_0}\delta_{{\bf m},
{\bf n}},
\end{eqnarray}
$K$ is the hopping parameter and the matrix $C$ is given in the next Section. The particular
time $N_0$ appearing depends on the fact that we inserted the projection
operator at the latest time, but the value of $ det \, Q|_{n_B}$ does not depend on
this arbitrary choice, as it will be clear from the derivation.

\section{Evaluation of the quark matrix in a given baryonic sector}

In the first part of this  Section we report the definition of the transfer matrix
$\hat{\cal{T}}
$  and the proof of Eq.~\reff{trace}. These results have been derived by
L\"uscher~\cite{Lusc}, and we will follow his paper even in the notation with minor changes.
Then we will evaluate the integral~\reff{convol} getting the quark matrix in a
given  baryonic sector.

To start with we report the explicit expression of the quark matrix
\be
Q= K \left[ (1+\gamma_0) U_0 T^{(+)}+ (1- \gamma_0) T^{(-)} U_0^+  +2 C \right] -B
\ee
where
\begin{eqnarray}
C &=& { 1 \over 2} \sum_{j=1}^3 
\left( U_j -  U_j^+ \right) \, \gamma_j,
\nonumber\\
B &= &1\!\!1 - K\sum_{j=1}^3 \left( U_j +  U_j^+ \right)
\end{eqnarray}
and the lattice spacing has been set equal to 1. The modifications necessary at 
nonzero temperature are obvious.
 The transfer matrix is defined in terms of the operator
 \begin{eqnarray}
 \hat{T}_F(n_0)  &= &  \exp\left( 2 K \hat{x}^+ B(n_0)^{-{ 1\over 2}} c(n_0)
B(n_0)^{-{ 1\over 2}}  \hat{y}^+ \right)
\nonumber\\
 & & \cdot \exp\left( -\hat{x}^+ M(n_0) \, \hat{x}  -\hat{y}^+ M(n_0) \, \hat{y} \right) 
\end{eqnarray}
according to the ordered product
\be
\hat{\cal{T}} = {\cal J}\prod_{n_0=-N_0}^{N_0-1} \left(\hat{T}_F(n_0)\right)^+
   \, \hat{T}_F(n_0+1). \label{transfm}
\ee
${\cal J}$ is the jacobian of a transformation which will be defined later, and
two new matrices have made their appearance \footnote{L\"uscher assumes the 
gauge $U_0= 1\!\!1$ to prove reflection positivity. We are not concerned here with
this most important issue and do not fix the gauge.}
\begin{eqnarray}
M &=& - \ln \left( (2K)^{{1 \over 2}} U_0^+ B^{-{ 1 \over 2}} \right)
\nonumber\\
c & = & { 1 \over 2} \sum_{j=1}^3 
\left( U_j -  U_j^+ \right) \, i \sigma_j.
\end{eqnarray} 
 The $\sigma_j$ are the Pauli matrices and the operator $ \hat{ T}_F(n_0) $  
depends on the time 
$n_0$ only throu the dependence on it of the gauge fields. 

We then use the following equations. Firstly, for arbitrary matrices $M$ and $N$
\be
<x|\exp \hat{ x}^+ M \, \hat{x} |x'> = \exp \left( x^+ \exp M \, x' \right),
\ee
 
\be
<x|\exp (\hat{x}^+ M  \, \hat{x}) \exp (\hat{x}^+ N \, \hat{x})|x'> = 
\exp \left( x^+ e^M e^N  x' \right). 
\ee
Secondly, if $B=B(\hat{x}^+)$ and $C = C( \hat{x})$ are operators which depend
on $ \hat{x}^+, \hat{x}$ only, for any operator A 
\be
<x|B \cdot A \cdot C |x'>= B( x^+) A( x^+,x') C(x').
\ee

We thus get the kernel 
\begin{eqnarray}
& &<x(n_0),y(n_0)| \hat{T}_F^+(n_0) \, \hat{T}_F(n_0+1)|x(n_0+1),y(n_0+1)> =
\nonumber\\
& & \,\,\,\,\,\,\,\,\,\,
\exp\left( 2 K x(n_0) B(n_0)^{-{ 1\over 2}} c(n_0) \,  B(n_0)^{-{ 1\over 2}} y(n_0) \right) 
\nonumber\\
 & & \,\,\,\,\,\,\,\,\,\,  \cdot   \exp\left( x^+(n_0)   e^{M(n_0)}  e^{M(n_0+1)} 
x(n_0+1)   - y^+(n_0)  e^{M(n_0)} e^{M(n_0+1)}  y(n_0+1) \right)
\nonumber\\
 & & \,\,\,\,\,\,\,\,\,\,  \cdot \exp\left( 2 K y(n_0+1) B(n_0+1)^{-{ 1\over 2}} c(n_0+1) \, 
B(n_0+1)^{-{ 1\over 2}} x(n_0+1)\right).  
\end{eqnarray}

Now to to go back from the Grassmann variables $x$' and $y$' to the quark field $\psi$ we
only need the relations 
\begin{eqnarray}
\psi &=&  B^{-{1 \over 2}} \pmatrix{x \cr y^+}
\nonumber\\
\overline{\psi} &=& \pmatrix{x^+  & y} \gamma_0 \,  B^{-{1 \over 2}}. \label{transf}
\end{eqnarray}
 This transformation cancels out the jacobian ${\cal J}$. Inserting the identity
between the factors in Eq.\reff{transfm}, using the equality
\be
\overline{\psi}_n \left( B \psi \right)_n = x_n^+ x_n + y_n^+y_n,
\ee
and collecting all the exponents we get Eq.~\reff{trace}.
At this point two observations are in order. The first is that the interpretation of
${\hat x},{\hat y}$ as annihilation operators of quarks and antiquarks follows from
the action of the charge conjugation on the quark field
\begin{eqnarray}
\psi' &=& {\cal C}^{-1} \, \overline{ \psi}
\nonumber\\
 \overline{\psi}' &=& -  \, \tilde{{\cal C}} \,\overline{ \psi},
\end{eqnarray}
where, with L\"uscher's convention for the $\gamma$-matrices,
\be
{\cal C}= \gamma_0 \gamma_2.
\ee
In fact these tranformations imply
\begin{eqnarray}
x' & = & -i \, \sigma_2 \,y
\nonumber\\
y' & = & i \,  \sigma_2 \, x.
\end{eqnarray}
The second observation is that all the above construction requires the Wilson 
parameter $r$ to be equal to 1.

There remains the evaluation of Eq.~\reff{convol}. We notice that 
\be
<x,y|\hat{\cal{T}}|z,w> = \tau \exp\left( x^+(N_0-1) D z + y^+(N_0-1)  D^* w + w E z
\right),
\ee
where $\tau$ does not depend on $ z,w$, the star means complex conjugation and
\begin{eqnarray}
D & = & B^{- {1\over2}}(N_0-1)\, U_0(N_0-1) \, B^{- {1\over2}}(N_0)
\nonumber\\
E & = &  B^{- {1\over2}}(N_0)  \, c (N_0)\, B^{- {1\over2}}(N_0).
\end{eqnarray}
We must therefore perform the gaussian integral
\begin{eqnarray}
& & <x,y|\hat{\cal{T}} P_{n_B}|-x,-y> =
\int [d z^+ dz \,dw^+ dw] \tau \exp \left( -z^+ z - w^+ w \right) \cdot
\nonumber\\
 & &\,\,\,\,\,\,\,\,\,\,    \exp \left( - j_1 z^+ x -j_2 \,  w^+ y +x^+(N_0-1)Dz +
 y^+(N_0-1) D^* w + w \, E \, z \right)
\nonumber\\
& & \,\,\,\,\,\,\,\,\,\, = \tau 
\exp \left( - j_1 x^+(N_0-1) D  \, x - j_2 \, y^+(N_0-1) D^* y +
j_1j_2\,y\,E\,x\right).
\end{eqnarray}
Comparing the kernel of the unconstrained transfer matrix with the above and taking into
account the antiperiodic boundary conditions of the quark field in euclidean time we get the
expression of Eq.~\reff{delta} for $\delta\, Q$.

\section{Conclusion}

In principle by the present approach one can
perform investigations of QCD in any definite baryonic sector at any
temperature, starting from baryon
number zero up to high baryonic densities. In practice, apart from the lowest 
baryonic numbers, we are
confined to small physical volumes. With the volumes actually accessible in
numerical simulations, perhaps
this is not a too severe limitation, and we can hope to get reasonable
results with the first few terms at least for not too high density and temperature. 

One could think of different applications of our expansion by further specifying
or changing the projection operator. 
In the case of baryonic number 1, for instance, by evaluating the expectation value
of the total angular momentum~\cite{X.Ji} one can disentangle the valence and cloud
contributions to the spin of the nucleon.
 Remaining to the lowest baryonic numbers, one
might try to attack the study of the deuteron. 

Another obvious field where our approach is potentially relevant is the QCD
phase diagram. In particular one could investigate the various superconductive
phases with spontaneous breaking of the color symmetry (see for instance~\cite{Wilc}) by
replacing the projection operator by the trial  state of the BCS theory.

\subsection*{Acknowledgements}

It is a pleasure to thank dr.G.DiCarlo for many conversations about QCD at
finite baryonic density.


\clearpage 

\begin{thebibliography}{9}


\bibitem{Math}
N.Mathur, S.J.Dong, K.F.Liu and N.C.Mukhopadhyay, Nucl.Phys B (Proc.Suppl.) 
83-84 (2000) 247; N.Mathur and S-J. Dong, Nucl. Phys. B (Proc.Suppl.) 94 (2000) 311;
for a review, see for instance H.Y.Cheng, Int. J. Mod, Phys. A11 (1996) 5109

\bibitem{Sand} 
 S.Sands, hep-lat/0109034, Sept.28, 2001

\bibitem{Mill}
D.E.Miller and K.Redlich, Phys. Rev. D35 (1987) 734;
A.Roberge and N.Weiss, Nucl. Phys. B275 [FS17] (1986) 734

\bibitem{Kacz}
O.Kaczmarek, with J.Engels, F.Karsh and E.Laermann, Nucl.Phys. B 558 (1999) 307;
Nucl.Phys. Proc. Suppl. 83 (2000) 369

\bibitem{Lusc}
M.L\"uscher, Commun. Math. Phys. 54 (1977) 283

\bibitem{Nege}
J.W.Negele and H.Orland, Quantum particle systems, Frontiers in Physics,
Addison-Wesley Publishing Company,1988

\bibitem{X.Ji}
X.Ji, Phys. Rev. Lett. 78 (1997) 610; 79 (1997) 1225


\bibitem{Wilc}
 F.Wilczek, hep-ph/0003183, 17 March 2000


\end{thebibliography}
\end{document}